\documentclass[journal]{IEEEtran}

\usepackage{cite}
\usepackage{amsmath,amssymb,amsfonts}
\usepackage{algorithmic}
\usepackage{graphicx}
\usepackage{textcomp}
\usepackage{url} 
\usepackage{fancyhdr}

\begin{document}

\setlength{\parindent}{5pt}
\setlength{\textfloatsep}{5pt}

\bstctlcite{IEEEexample:BSTcontrol}
\title{A Fused Deep Denoising Sound Coding Strategy for Bilateral Cochlear Implants}
\author{\emph{Tom Gajecki and Waldo Nogueira}
\thanks{TG, and WN are with the Department of Otolaryngology, the Medical University of Hannover and Cluster of Excellence, Hearing4all, Hannover, 30625,
Germany. The funding for this work was provided by the German Research
Foundation (DFG) under Project ID 446611346, led by Waldo Nogueira.}}

\maketitle
\thispagestyle{fancy}

\begin{abstract}
Cochlear implants (CIs) provide a solution for individuals with severe sensorineural hearing loss to regain their hearing abilities. When someone experiences this form of hearing impairment in both ears, they may be equipped with two separate CI devices, which will typically further improve the CI benefits. This spatial hearing is particularly crucial when tackling the challenge of understanding speech in noisy environments, a common issue CI users face. Currently, extensive research is dedicated to developing algorithms that can autonomously filter out undesired background noises from desired speech signals.
At present, some research focuses on achieving end-to-end denoising, either as an integral component of the initial CI signal processing or by fully integrating the denoising process into the CI sound coding strategy. This work is presented in the context of bilateral CI (BiCI) systems, where we propose a deep-learning-based bilateral speech enhancement model that shares information between both hearing sides. Specifically, we connect two monaural end-to-end deep denoising sound coding techniques through intermediary latent fusion layers. These layers amalgamate the latent representations generated by these techniques by multiplying them together, resulting in an enhanced ability to reduce noise and improve learning generalization. The objective instrumental results demonstrate that the proposed fused BiCI sound coding strategy achieves higher interaural coherence, superior noise reduction, and enhanced predicted speech intelligibility scores compared to the baseline methods. Furthermore, our speech-in-noise intelligibility results in BiCI users reveal that the deep denoising sound coding strategy can attain scores similar to those achieved in quiet conditions.
\end{abstract}


\section{Introduction}

\IEEEPARstart{A}{}cochlear implant (CI) is a medical device surgically implanted to restore the sense of hearing in individuals with severe to profound sensorineural hearing loss. Notably, recent years have seen significant advancements in CI technology \cite{Lenarz2018}. Consequently, individuals with bilateral hearing loss often receive implants on both sides \cite{kan2015}. Those who receive a CI in each ear are known as bilateral CI (BiCI) users, typically demonstrate improved speech understanding, sound localization, reduced listening effort, and enhanced quality of life in comparison to unilateral CI users (e.g., \cite{litovsky2006c, asp2012, hughes2013, Schoonhoven2013}). However, their listening performance remains inferior to individuals with normal hearing (NH) (e.g., \cite{Loizou2009, murphy2011, Kerber2012}).

The disparity in performance between BiCI users and individuals with NH could potentially stem from differences in electrode array insertion depth, differences between the electrode-nerve interfaces in each ear, and from the independent processing in each CI (e.g., \cite{hoesel2003, dennison2022impact, gajecki02, gajecki03}). The computation of stimulation current levels over time and for individual electrodes (referred to as electrodograms) relies on audio captured by microphones embedded within each speech processor. This computation is achieved by applying the CI sound coding strategy independently to both listening sides, which can lead to a lack of effective binaural integration \cite{kan2013_2}, may introduce binaural artifacts \cite{goup2013}, may struggle to effectively suppress background noise or competing speech signals when present in both ears simultaneously \cite{gajecki02}, and might not fully transmit interaural cues \cite{Williges2018}.

Typically, a CI in conjunction with its associated sound coding strategy enables the user to understand speech effectively in quiet environments. However, its effectiveness diminishes when encountering loud interfering signals, characterized by low signal-to-noise ratios (SNRs), such as background noise or multiple speakers talking simultaneously \cite{Hochberg}. Several approaches have been proposed to enhance speech understanding in noisy environments for BiCIs. Some of these methods utilize traditional front-end processing techniques like binaural beamforming (e.g., \cite{guevara2016voice, tammen2022deep, baumgartel2015comparing}), while others integrate elements of the CI sound coding strategy and establish bilateral connections between certain processing components (e.g., \cite{Lopez-Poveda2016,lopez2017, gajecki02}). These conventional approaches have proven effective in augmenting speech understanding in noise and sound source localization for BiCI users. However, with the advent of deep learning technology, the field is increasingly exploring the use of deep neural networks (DNNs) for speech enhancement (e.g., \cite{luo2019conv, lai2016deep,Lai2018, hu2010,mamun2019convolutional}). These methods have proven to be very successful at performing speech denoising while keeping speech quality and a high degree of generalization capabilities.

To optimize the enhancement of speech for CIs, it could prove advantageous to devise algorithms that take into account the specific processing scheme of CIs. Consequently, there has been research dedicated to CIs, where DNNs are incorporated into their signal pathway \cite{bolner2016, nogueira2016-2, goehring2017, mamun2019convolutional, zheng2021noise}. These approaches target noise reduction by directly applying masks within the filter bank utilized by the CI sound coding strategy. Recently, drawing inspiration from the Conv-TasNet \cite{luo2019conv}, an end-to-end CI sound coding strategy based on deep learning, termed ``Deep ACE.'' was proposed \cite{gajecki2022end, gajecki2023deep}. This approach was designed to replace the clinically available ACE sound coding strategy, but it could also be used to replace other commercially available ones. This method completely replaces the CI sound coding strategy with a DNN and achieves high speech understanding improvements in BiCI users (up to 22.8\% improvement in word recognition score (WRS) in modulated background noise).

Presently, data-driven methodologies have primarily been employed with single CIs. However, these approaches are equally applicable to BiCIs. For instance, there is no inherent justification to assume that utilizing any monaural speech enhancement algorithm within a BiCI framework would not produce comparable auditory advantages as observed in the unilateral configuration. Nevertheless, this may not be the optimal approach, as independent processing could still potentially introduce artifacts that hinder effective binaural listening. A promising avenue to enhance the listening experience of BiCI users involves embracing multi-channel sound processing. Notably, diverse multi-channel front-end speech enhancement methods have been proposed. These strategies not only showcase effectiveness in enhancing speech denoising but also reveal an ability to maintain the integrity of essential binaural auditory cues (e.g., \cite{Han2020, 9053092}).

In recent advancements, there's a novel concept referred to as ``Fusion Layers'' introduced in \cite{gajecki04}. These layers entail the exchange of information between two individual monaural speech denoising algorithms. They achieve this by enabling Hadamard products between latent spaces at specific processing stages, drawing inspiration from multi-task learning methods, and emulating the inhibitory and excitatory mechanisms found in the human brain stem for binaural hearing \cite{moore1991}. Precisely, the fusion layers are designed to introduce non-linear elements into the learning model, enhancing the model's ability to fit training data effectively while improving generalization without impacting the number of trainable parameters. This approach of sharing features has proven to be highly effective in enhancing noise reduction compared to independent bilateral models, where processing is performed separately on each side.

In our study, we present a novel approach termed the ``fused Deep ACE,'' which can naturally be extrapolated to other CI processing strategies. This approach integrates two Deep ACE algorithms through the utilization of fusion layers, enabling the sharing of latent representations from particular processing stages. Our hypothesis is that this bilateral sound coding strategy will result in improved speech understanding when contrasted with the conventional clinical approach. Additionally, we postulate that the fusion layer will capitalize on bilateral redundant information, potentially mitigating certain binaural artifacts and leading to the generation of more bilaterally coherent output electrodograms.

\section{Methods \& Materials}
\subsection{Bilateral advanced combination encoder (ACE; unprocessed)}
This is the main baseline algorithm used in this work and is based on a clinical BiCI setup, where each CI processes the sound independently using the ACE sound coding strategy. This setup does not perform any noise reduction and does not share any information between the listening sides. The ACE strategy begins by sampling the acoustic signal at 16 kHz, followed by applying a filter bank through a 128-point fast Fourier transform. This process introduces a 2 ms algorithmic latency, dependent on the channel stimulation rate (CSR). Estimations of desired envelopes are calculated for each spectral band ($E_k$) corresponding to an electrode, with $M$ representing the total channels.

In this study, we select the $N$ most energetic envelopes out of $M$ based on their amplitudes. These selected envelopes undergo non-linear compression via a loudness growth function (LGF). The LGF output ($p_k$) represents the normalized stimulation amplitude for electrode $k$ to stimulate the auditory nerve. Lastly, we map each $p_{k}$ within the subject's dynamic range, spanning from threshold to comfortable stimulation levels giving the output current stimulation patterns $I_k$. These $N$-selected electrodes are stimulated sequentially for each audio frame, defining one stimulation cycle, and the CSR is determined by the cycles per second.

\subsubsection{Bilateral Deep ACE} 
This condition closely resembles the baseline scenario (referred to as bilateral ACE) in that it lacks any exchange of information between the listening sides. However, it diverges from clinical ACE sound coding strategies by adopting the newly developed Deep ACE approach, as detailed in \cite{gajecki2022end, gajecki2023deep}. More specifically, Deep ACE substitutes the conventional clinical ACE method with a DNN that takes in raw audio as its input and generates the denoised LGF output $p_k$.

In the initial step, Deep ACE encodes the left and right signals $X_{\{l,r\}}$ into a latent representation using a 1-D convolution layer. This operation can be mathematically expressed as a matrix multiplication:
\begin{equation}
    X'_{\{l,r\}}=\Theta(X_{\{l,r\}}\cdot {\mathrm{\mathbf{E}}}),
\end{equation}
where ${\mathrm{\mathbf{E}}}_{\{l,r\}} \in \Re^{(F\times L)}$ are the left and right encoder basis functions and $\Theta(\cdot)$ is the antirectifier activation function used in Deep ACE, and $F$ and $L$ the number and length (in samples) of the filters used, respectively. 
The signal is then sent to a deep envelope detector (DED) that performs dimensionality reduction (from $F$ to $M$) and to the separator module that will generate a deeper latent representation for each side $X''_{l,r}=\zeta(X'_{\{l,r\}}) \in \Re^{(1\times S)}$, where $\zeta(\cdot)$ is the learned function by the separator and $S$ is the number of skip connections \cite{luo2019conv}. Then the DED and separator outputs are fed into a masker that will remove the noisy components of the encoded mixture. Finally, the masked signals will be decoded through a transposed 1-D convolution to obtain $p_k$ for each CI.

\subsubsection{Fused Deep ACE} 
\begin{figure}
	\centering
	\includegraphics[width = .48\textwidth]{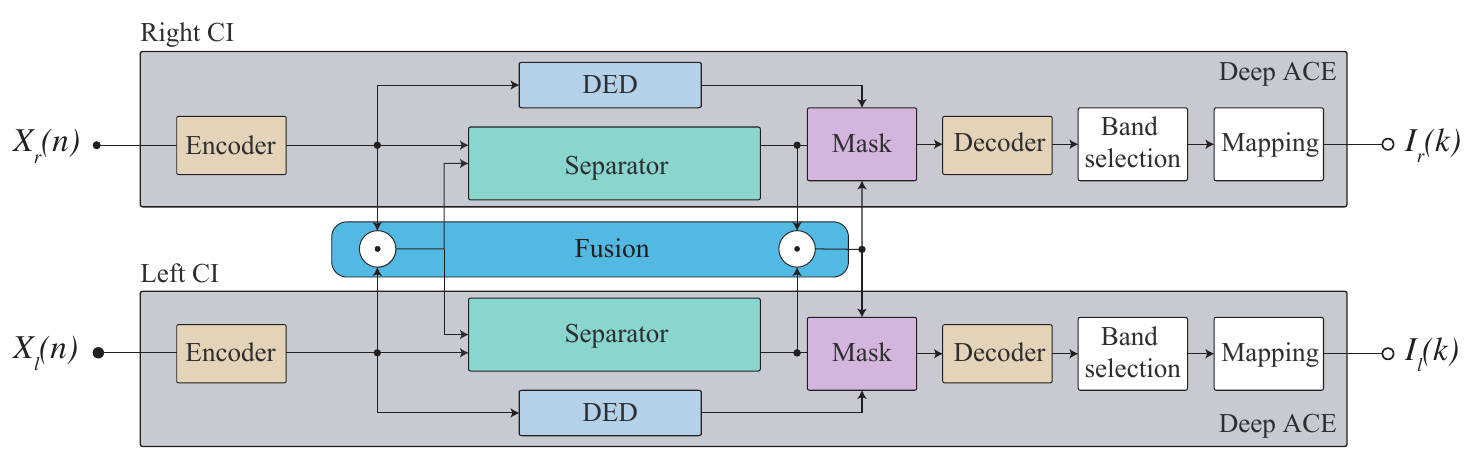}  
    \caption{Block diagram of the proposed fused Deep ACE. The model takes the right and left time-domain noisy speech signals ($X_r(n)$ and $X_l(n)$ respectively) and produces the respective denoised current stimulation patterns for each listening side $I_r(k)$ and $I_l(k)$ for each stimulation frame $k$. The fusion model performs element-wise dot product between the latent representations generated in each of the Deep ACE models and the deep envelope detector (DED) is used for dimensionality reduction.}
\label{bin_deep_Ace}
\end{figure}

In this study, we introduce an approach involving integrating two monaural Deep ACE \cite{gajecki2023deep} models, with one model associated with each listening side. This integration is achieved through the utilization of fusion layers \cite{gajecki04}. These fusion layers are influenced by the principles of multi-task learning, where model weights are shared across different models to address interconnected tasks. The function of these layers involves conducting element-wise dot products on tensors that depict latent representations at identical processing stages. More precisely, we combine the latent representations generated within each Deep ACE model, both following the encoding stage and subsequent to the separator modules as follows:

\begin{equation}
 \begin{aligned}
    X'_\Lambda = \rho(X'_l, X'_r) \\
    X''_\Lambda = \rho(X''_l, X''_r),
\end{aligned}
\end{equation}
where $\rho(\cdot)$ is the Hadamard product operator. The outcome of these two fusion operations results in a model that performs ``double fusion.'' These fused signals are fed into the separator and masker modules the same way as in the bilateral Deep ACE condition.
A visual representation of this model's structure can be observed in Figure \ref{bin_deep_Ace}. It is important to note that, within this model, the functions of the band selection and mapping blocks are mutually exclusive and unaffected by one another.

\subsection{Model training setup}
Training the models was conducted over a maximum of 100 epochs, employing batches consisting of two 4-second-long audio segments. The initial learning rate was initialized to 1e-3. In case the validation set accuracy displayed no enhancement over a span of 3 consecutive epochs, the learning rate was reduced by half. To ensure regularization, early stopping with a patience of 5 epochs was implemented, safeguarding against overfitting. Only the model displaying the highest performance was retained. Model optimization was facilitated using the Adam optimizer \cite{adam}. The model's training employs a cost function based on mean-squared error (MSE) and binary cross entropy (BCE) for each listening side (for a detailed description of this cost function refer to \cite{gajecki2023deep}).

The hyperparameter configuration used in this study was slightly modified with respect to the ones shown in \cite{gajecki2022end}, specifically the separator module was bigger and the deep-envelope-detector (DED) was also increased in size. The model hyperparameters are shown in Table \ref{hyperparameters}.

\begin{table}[h!]
\caption{Hyperparameters used to train the deep learning models. The tuple corresponding to the DED value shows the number of channels contained in each of the convolutional layers}
\label{hyperparameters}
\centering
\begin{tabular}{lcc}
\hline
Parameter && Value    \\ \hline 
\hline
F             && 64         \\ \hline
L             && 32          \\ \hline
B             && 64           \\ \hline
H             && 128      \\ \hline
S             && 32            \\ \hline
P             && 3             \\ \hline
X             && 8               \\ \hline
R             && 3              \\ \hline
DED            && (128,64,$M$)              \\ \hline
\end{tabular}
\end{table}

\subsection{Audio material}
\label{audio_data}
In this work, we used a total of three different speech datasets and three noise types to assess the models' performance and generalization abilities. All these audio sets will be described in this section. As a preprocessing stage, all audio material was set to mono and resampled at 16 kHz. The corresponding electrodograms were obtained by processing all audio data with the ACE sound coding strategy at an output CSR of 1,000 pulses per second. All audio signals were generated by convolving source signals with binaural room impulse responses (BRIRs; \cite{brirs}) and summing. BRIRs were generated for hearing aids located in each listening side \footnote{\url{https://github.com/IoSR-Surrey/RealRoomBRIRs}} and consisted of 4 different rooms of different sizes and acoustic properties, using the front microphone. 

\subsubsection{Speech data}

\paragraph{\textbf{LibriVox corpus} \cite{beilharz2020}}  
This speech data was originally designed for end-to-end speech translation, however, in this study, we mix the speech material with noise to train our models for speech denoising. The speech data consists of fluent spoken sentences with a total duration of 18 hours. The quality of audio and sentence alignments was checked by a manual evaluation, showing that speech alignment is in general very high. In fact, the sentence alignment quality is comparable to well-used parallel translation data.

\paragraph{\textbf{TIMIT corpus} \cite{zue1990}} This corpus contains broadband recordings of 630 people speaking the eight major dialects of American English, each reading ten phonetically rich sentences. In this work, files from 112 male and 56 female speakers in the test set were selected.

\paragraph{\textbf{HSM corpus}\cite{hsm}}  Speech intelligibility in quiet and in noise was measured by means of the
Hochmair, Schulz, Moser (HSM) sentence test, based on a dataset composed of 30 lists with 20 everyday sentences each (106 words per list).
  
\subsubsection{Noise data}
\paragraph{\textbf{Environmental noises}; DEMAND \cite{demand}} The environmental noises recorded to create this dataset are split into six categories; four are indoor noises and the other two are outdoor recordings. The indoor environments are further divided into domestic, office, public, and transportation; the open-air environments are divided into streets and nature. There are 3 environment recordings per category.

\paragraph{\textbf{Synthetic noises}; SSN \cite{hugo2007} \emph{and ICRA7} \cite{icra}} To evaluate the different algorithms, in this work we also use stationary speech-shaped noise (SSN) and non-stationary modulated seven-speaker babble noise (ICRA7) as synthetic interferers.

\subsubsection{Training, evaluation and testing data}

The training set was composed of speech from the LibriVox corpus and noise from the DEMAND dataset. Specifically, 30 male (M) and female (F) speakers were randomly selected from the speech corpus, and two environments were randomly selected from each of the noise categories. For validation, 20\% of the training data was used. The noise and speech subsets used for training will be referred to as EN$_1$ and LibriVox$_1$, respectively. The testing phase involved the utilization of the HSM speech dataset, coupled with synthetic noises employed as interfering signals.

Speech and noise signals were mixed at SNR values ranging uniformly from -5 to 10 dB. The processed clean speech signals were also included in the listening experiments to assess whether the proposed model introduced perceptually relevant distortions. 

\subsection{Objective Evaluation}
To objectively evaluate the performance of each examined algorithm, we gauge the extent of noise reduction accomplished, establish electrode-wise correlation coefficients between the denoised and clean signals, and determine speech intelligibility through the application of the modified binaural short-time objective intelligibility (MBSTOI) index \cite{mbstoi}. Notably, in this study, our focus is on investigating comprehensive CI processing, consequently prompting the computation of the MBSTOI index from synthesized electrodograms ($\boldsymbol{p}$) derived using a vocoder. This results in the utilization of a specific variant of MBSTOI referred to as vocoder-MBSTOI (V-MBSTOI).

\subsubsection{SNRi}
To assess the amount of noise reduction performed by each of the tested algorithms we compute the SNR improvement (SNRi).
This measure is calculated in the electrodogram domain and compares the original input SNR to the one obtained after denoising, and is given by:
\begin{equation}
\mathrm {SNRi}=10 \cdot \log_{10}\Bigg(\frac{\sum_{k=1}^M||\boldsymbol{p}_k^{n}-\boldsymbol{p}_k^{c}||^2}{\sum_{k=1}^M||\boldsymbol{p}_k^{d}-\boldsymbol{p}_k^{c}||^2}\Bigg),
\label{snri}
\end{equation}
where $\boldsymbol{p}_k$ represents the LGF output of band $k$ and the superscripts $n$, $c$, and $d$ are used to denote the noisy, clean, and denoised electrodograms, respectively.

\paragraph{\textbf{V-MBSTOI}} 
To estimate the speech intelligibility performance expected from each of the algorithms, the V-MBSTOI score \cite{taal2010short, hinrich2021, watkins2018evaluation} was used. This metric relies directly on MBSTOI \cite{mbstoi}, which is modeled based on normal hearing binaural speech performance. Specifically, the purpose of this metric is to evaluate the potential relative variations in speech performance that could be achieved in behavioral experiments, rather than providing an exact estimation of an individual's CI performance. The V-MBSTOI score ranges from 0 to 1, where the higher score represents a predicted higher speech performance. 

\paragraph{\textbf{Linear cross-correlation}}
To characterize potential distortions and artifacts introduced by the tested algorithms, the linear correlation coefficients (LCCs) between the clean ACE electrodograms ($\boldsymbol{p}^c$) and the denoised electrodograms ($\boldsymbol{p}^d$) were computed. The LCCs were first computed channel-wise (i.e., one correlation coefficient was computed for each of the 22 channels) to assess channel output degradation caused by the denoising process. The $\mathrm{LCC}_k$ for band $k$ is computed based on the Pearson correlation coefficient \cite{freedman2007statistics} as follows:

\begin{equation}
\mathrm{LCC}_k=\frac{\operatorname{cov}\left(\boldsymbol{p}^{c}_k, \boldsymbol{p}_k^{d}\right)}{\sigma_{\boldsymbol{p}_k^{c}} \cdot \sigma_{\boldsymbol{p}_k^{d}}},
\label{lcc}
\end{equation}
where $\operatorname{cov}(X, Y)$ is the covariance between $X$ and $Y$, and $\sigma_{\boldsymbol{p}_k}$ is the standard deviation of the values in the corresponding electrodogram $\boldsymbol{p}_k$.

We also present the LCCs as a function of the noise azimuth $LCC_\theta$, which is computed as follows:

\begin{equation}
\mathrm{LCC}_\theta=\frac{\operatorname{cov}\left(\boldsymbol{p}^{c}_\theta, \boldsymbol{p}_\theta^{d}\right)}{\sigma_{\boldsymbol{p}_\theta^{c}} \cdot \sigma_{\boldsymbol{p}_\theta^{d}}},
\label{lcctheta}
\end{equation}
where $\boldsymbol{p}^{c}_\theta$ and $\boldsymbol{p}^{d}_\theta$ are the LCCs averaged across electrodes for a noise source coming from azimuth $\theta$.

\paragraph{\textbf{Electric interaural coherence}}
\label{eic}
Similar to the LCCs, we also use the electric interaural coherence (EIC). Here we compute the channel-wise LCCs between between the right electrodograms ($\boldsymbol{p}^r$) and the left electrodograms ($\boldsymbol{p}^l$) as follows:

\begin{equation}
\mathrm{EIC}_k=\frac{\operatorname{cov}\left(\boldsymbol{p}^{r}_k, \boldsymbol{p}_k^{l}\right)}{\sigma_{\boldsymbol{p}_k^{r}} \cdot \sigma_{\boldsymbol{p}_k^{l}}},
\label{icck}
\end{equation}

We also present the EIC as a function of the noise azimuth LCC$_\theta$, which is computed as follows:

\begin{equation}
\mathrm{EIC}_\theta=\frac{\operatorname{cov}\left(\boldsymbol{p}^{r}_\theta, \boldsymbol{p}_\theta^{l}\right)}{\sigma_{\boldsymbol{p}_\theta^{r}} \cdot \sigma_{\boldsymbol{p}_\theta^{l}}},
\label{icctheta}
\end{equation}
where $\boldsymbol{p}^{r}_\theta$ and $\boldsymbol{p}^{r}_\theta$ are the EIC averaged across electrodes for a noise source coming from azimuth $\theta$.

\begin{table}
\caption{CI participant information and experiment settings}
\label{Participants}
\centering
\begin{tabular}{lcccccc} 
\hline
ID     & Age  & Gender & $N$-of-$M$ & CCITT SNR  & ICRA7 SNR \\ \hline\hline
BI01   & 71   & M      & $8$-of-$20$   & 5 dB              & 5 dB   \\ \hline
BI02   & 72   & F      & $5$-of-$22$       & 0 dB               & 5 dB    \\ \hline
BI03   & 60   & F      & $8$-of-$19$         & 0 dB               & 0 dB    \\ \hline
BI04   & 70   & M      & $8$-of-$19$        & 0 dB               & 0 dB    \\ \hline
BI05   & 75   & M      & $8$-of-$22$       & -5 dB               & -5 dB    \\ \hline
\end{tabular}
\end{table}

\subsection{Behavioral evaluation}
To validate the objective instrumental measures and to assess their impact on actual BiCI hearing, we perform two behavioral experiments namely, a speech intelligibility experiment and a  Multiple Stimuli with Hidden Reference and Anchor (MUSHRA). The speech intelligibility experiments are designed to investigate the benefits of the proposed denoising algorithms when compared to the clinical setups, and the MUSHRA \cite{liebetrau2014revision} will help understand how BiCIs rate the quality of the performed denoising.

The stereo signals were transmitted through direct stimulation using a bilaterally synchronized RF GeneratorXS interface from Cochlear Ltd. (Sydney, Australia) in conjunction with MATLAB software (Mathworks, Natick, MA) via the Nucleus Implant Communicator V.3, also from Cochlear Ltd. All testing procedures were conducted on a personal computer equipped with customized MATLAB software. Prior to commencing experiments involving subjects, a hardware check was carried out by analyzing the signals generated by the research interface using an oscilloscope. The stimulation signals were characterized by cathodic-phase leading, biphasic pulses presented in a monopolar configuration (MP1+2). This stimulation mode utilizes two extracochlear electrodes: one ball electrode positioned under the temporalis muscle and another plate electrode on the receiver case. These pulses consistently featured an 8-$\mu$s phase gap and 25-$\mu$s phase duration, and they were presented in a base-to-apex sequence.

\subsubsection{Speech understanding experiment}

Speech intelligibility in noisy environments was assessed using the HSM sentence set \cite{hsm}. To conduct this assessment, each speech token underwent digital downsampling from 44.1 kHz to 16 kHz. During testing, subjects were presented with sentences from the front in a simulated acoustic setting, which included background interference noise (either CCITT or ICRA7) originating from a 55-degree azimuth angle, masking their self-reported better ear. The noise azimuth was selected to be 55 degrees because this angle corresponded to the point where electrical interaural coherence (EIC; described in \ref{eic}) was at its minimum (see Figure \ref{eic_azimuth}), thus maximizing the impact on speech understanding.

Before the speech tests began, a training phase was implemented, comprising two sets of 20 sentences presented in quiet conditions. This training allowed listeners to adapt to the fitting parameters specific to the study and familiarize themselves with the sound delivery through the research interface.

Subjects were instructed to verbally repeat the sentences as accurately as possible during the tests. Two observers were present during the tests: one managed the software interface, while the other recorded the number of correctly identified words by marking them in a printed list corresponding to the sentences. Each listening condition was evaluated twice using different sentence lists, and the final score was computed as the average number of correctly identified words across these repetitions. The subjects were unaware of the specific conditions being tested, and an audiologist, blind to the test conditions, conducted the speech intelligibility assessments.

\subsubsection{MUSHRA}
This test is aimed at assessing how well-presented speech sentences are perceived in comparison to a specified reference using MUSHRA. The scores provided by the listener will range from 0 (poor) to 100 (excellent). In the context of this study, the primary goal of this experiment was to establish a relative score for the quality of speech denoising concerning the clean speech signal generated by the clinical sound coding strategy ACE. To create a reference point, we derived an anchor by applying a low-pass filter with a cut-off frequency of 3.5 kHz to the noisy, unprocessed mixture. Two primary conditions were examined: one with clean audio and the other in a noisy environment (using both CCITT and ICRA7 noise profiles).

In the clean condition, we compared the reference clean ACE to the anchor and the clean speech signals processed separately by the independent BiCI strategy and the fused Deep ACE  sound coding strategy. This comparison aimed to determine if there were discernible differences between clinical processing in a quiet setting and the proposed algorithms.

In the noisy condition, we compared the reference clean ACE to the anchor, the two proposed algorithms, and the unprocessed ACE signal in a noisy environment. Within each MUSHRA block, corresponding to each primary condition, eight sentences were assessed. These sentences were presented at various SNRs, including 2 at -5dB, 2 at 0dB, 2 at 5dB, and 2 at 10 dB.

\section{Results}
\subsection{Objective instrumental results}
\begin{figure}[h!]
	\centering
	\includegraphics[width = .48\textwidth]{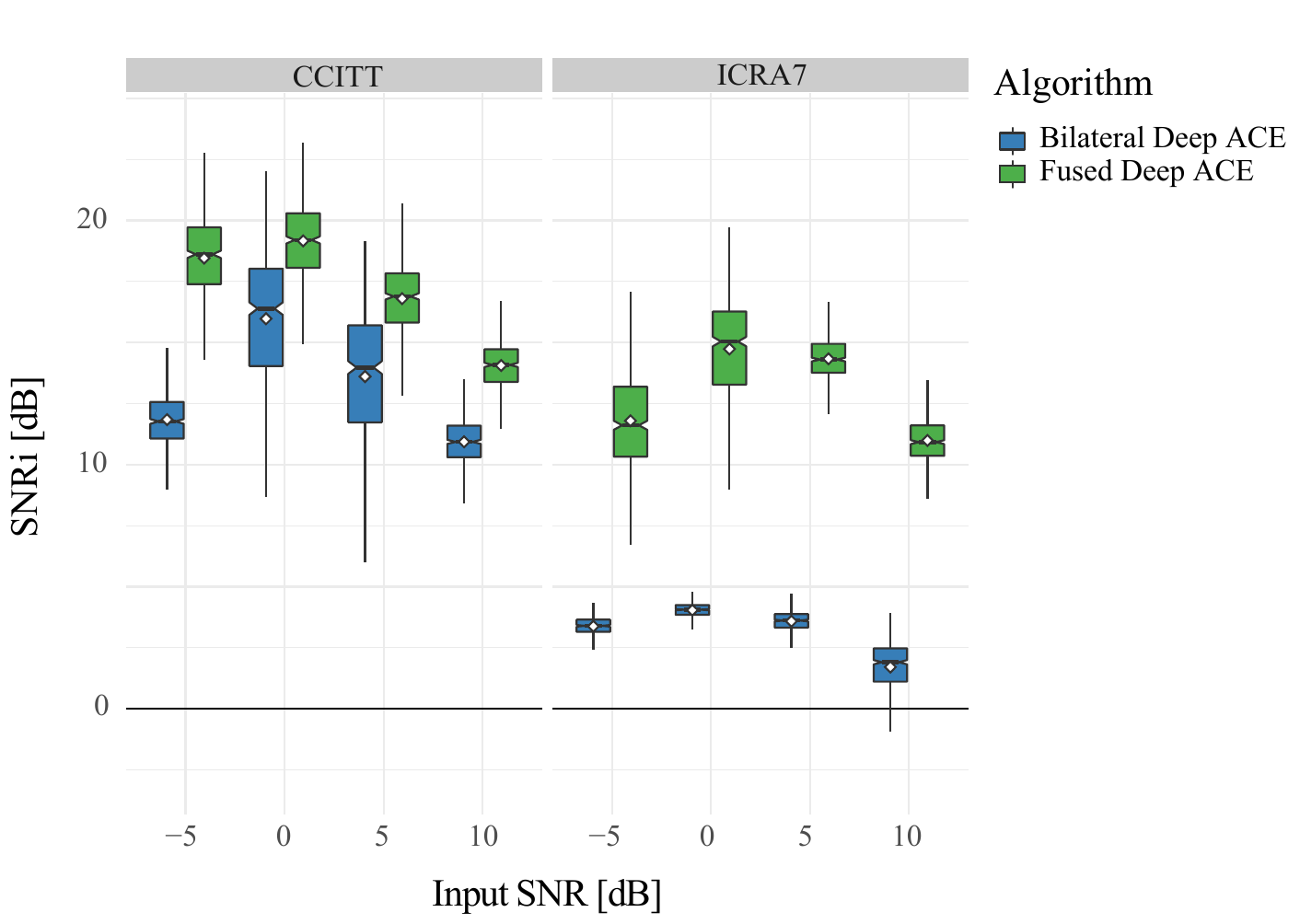}  
    \caption{Box plots showing the mean SNRi scores across listening sides in dB for the tested algorithms in CCITT and ICRA7 noises for the different SNRs using the HSM speech dataset. The black horizontal bars within each box represent the median for each condition, the diamond-shaped marks indicate the mean, and the top and bottom extremes of the boxes indicate the $Q_3=75\%$ and $Q_1 = 25\%$ quartiles, respectively. The box length is given by the interquartile range (IQR), used to define the whiskers that show the variability of the data above the upper and lower quartiles (the upper whisker is given by $Q_{3} + 1.5\cdot$IQR and the lower whisker is given by $Q_{1} - 1.5\cdot$IQR \cite{rstudio})}
    \label{snri_fig}
\end{figure}

\paragraph{\textbf{SNRi}} Figure \ref{snri_fig} shows box plots showing the mean SNRi scores across listening sides in dB for the tested algorithms in CCITT and ICRA7 noises for the different SNRs using the HSM speech dataset.

\begin{figure}[h!]
	\centering
	\includegraphics[width = .48\textwidth]{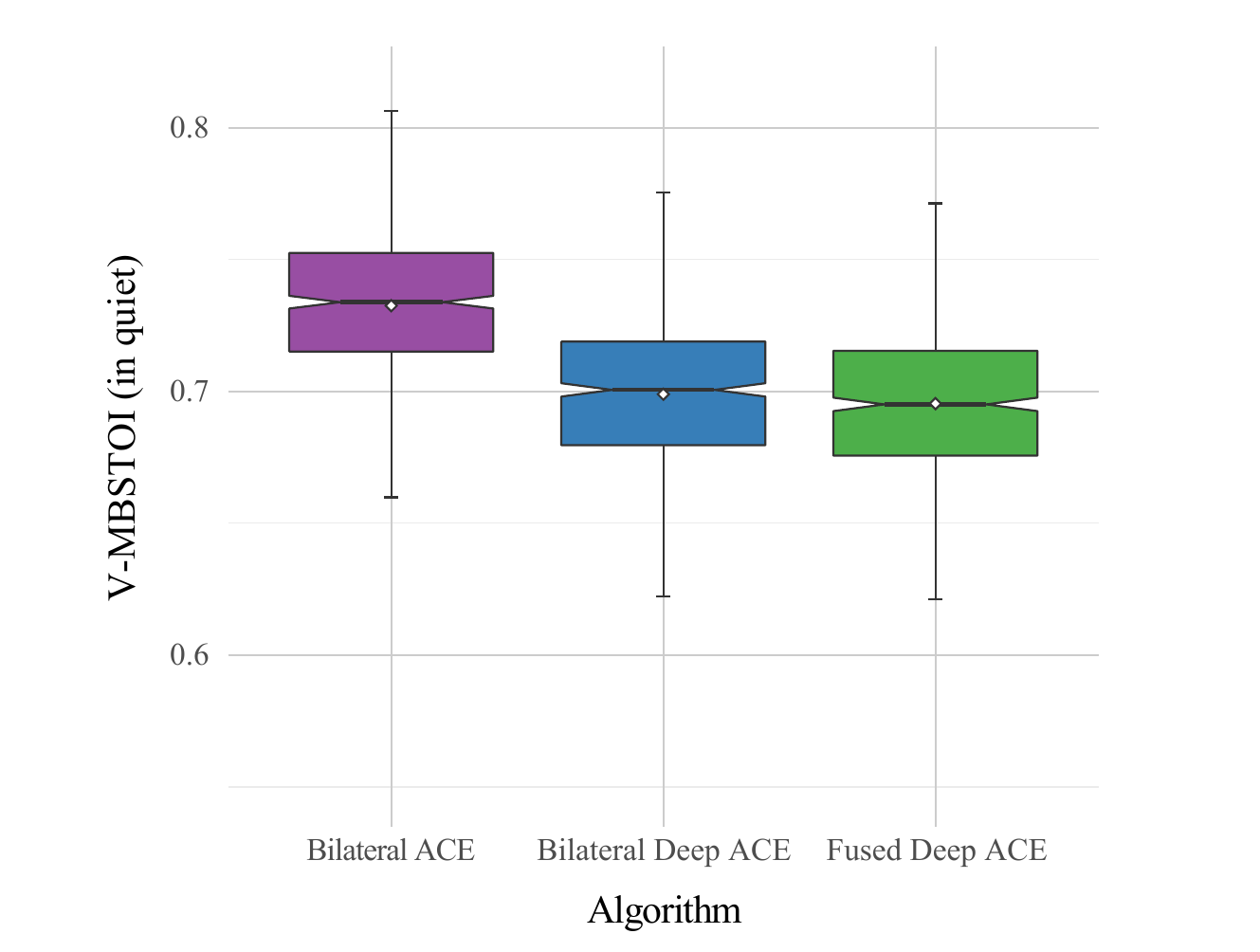}  
    \caption{Box plots showing the V-MBSTOI scores for the tested algorithms in quiet for the different SNRs using the HSM speech dataset. The black horizontal bars within each box represent the median for each condition, the diamond-shaped marks indicate the mean, and the top and bottom extremes of the boxes indicate the $Q_3=75\%$ and $Q_1 = 25\%$ quartiles, respectively. The box length is given by the interquartile range (IQR), used to define the whiskers that show the variability of the data above the upper and lower quartiles (the upper whisker is given by $Q_{3} + 1.5\cdot$IQR and the lower whisker is given by $Q_{1} - 1.5\cdot$IQR \cite{rstudio}).}
    \label{mbstoi_clean}
\end{figure}

\paragraph{\textbf{V-MBSTOI}}
 Figure \ref{mbstoi_clean} illustrates the V-MBSTOI scores obtained by the evaluated algorithms in quiet. It can be seen here that the denoising algorithms do not introduce a significant drop in the V-MBSTOI scores relative to the bilateral ACE condition.

\begin{figure}[h!]
	\centering
	\includegraphics[width = .48\textwidth]{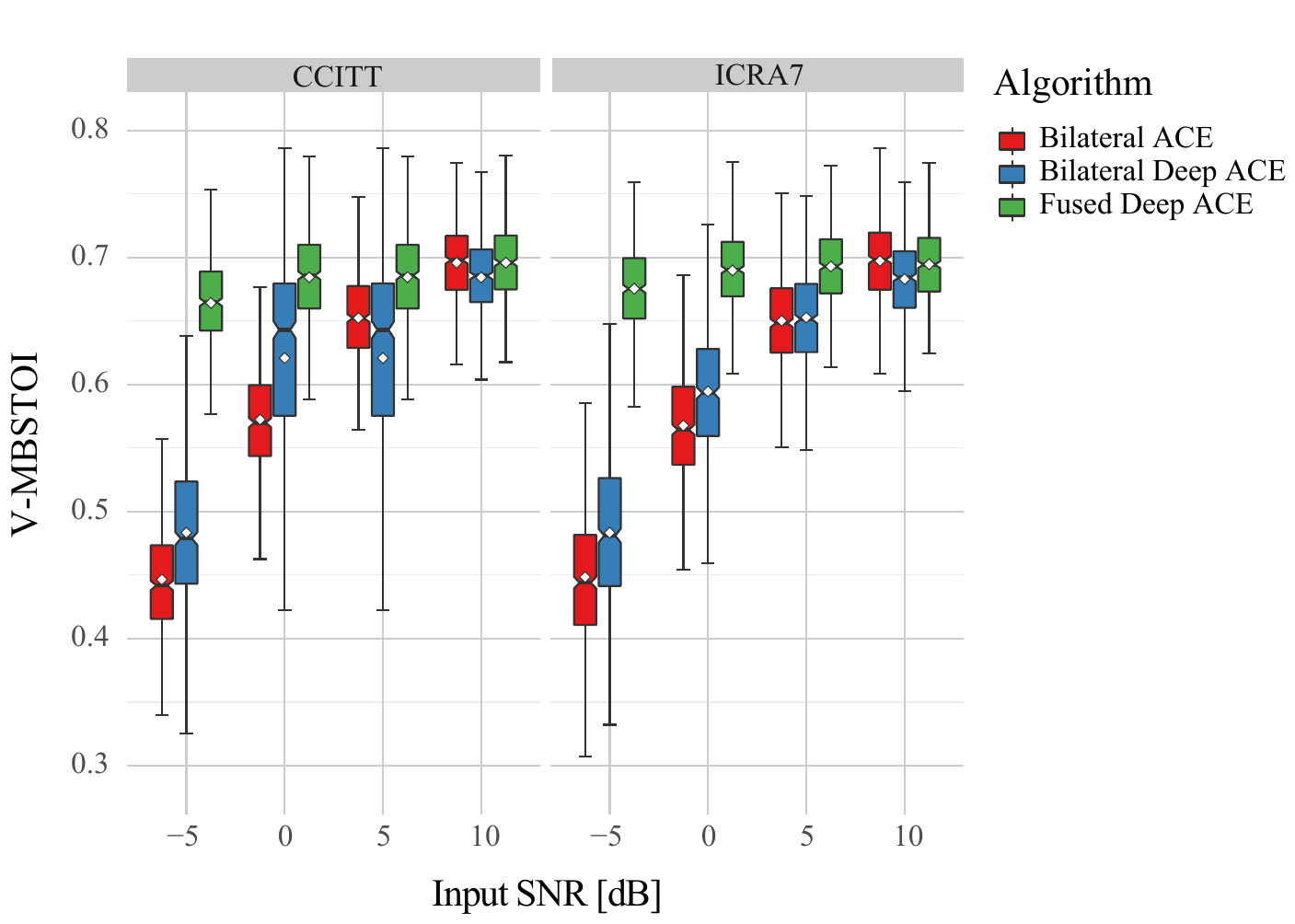}  
    \caption{Box plots showing the V-MBSTOI scores for the tested algorithms in CCITT and ICRA7 noises for the different SNRs using the HSM speech dataset. The black horizontal bars within each box represent the median for each condition, the diamond-shaped marks indicate the mean, and the top and bottom extremes of the boxes indicate the $Q_3=75\%$ and $Q_1 = 25\%$ quartiles, respectively. The box length is given by the interquartile range (IQR), used to define the whiskers that show the variability of the data above the upper and lower quartiles (the upper whisker is given by $Q_{3} + 1.5\cdot$IQR and the lower whisker is given by $Q_{1} - 1.5\cdot$IQR \cite{rstudio}).}
    \label{mbstoi_noise}
\end{figure}

Figure \ref{mbstoi_noise} presents the V-MBSTOI scores achieved by the assessed algorithms under different speech and noise conditions. Generally, the denoised signals exhibit higher scores using the denoising algorithms compared to the bilateral ACE, and the improvement is roughly proportional to the input SNR (calculated at the better SNR side).

However, it is noteworthy that the bilateral Deep ACE model falls short of the fused speech denoising method, indicating that the artifacts in the latter are comparatively smaller. Additionally, the V-MBSTOI scores computed across various input SNRs exhibit less variability for the fused Deep ACE model when compared to the bilateral Deep ACE and bilateral ACE counterparts. This suggests that the fused Deep ACE model may demonstrate greater robustness in scenarios with low input SNRs.

\begin{figure}[h!]
	\centering
	\includegraphics[width = .48\textwidth]{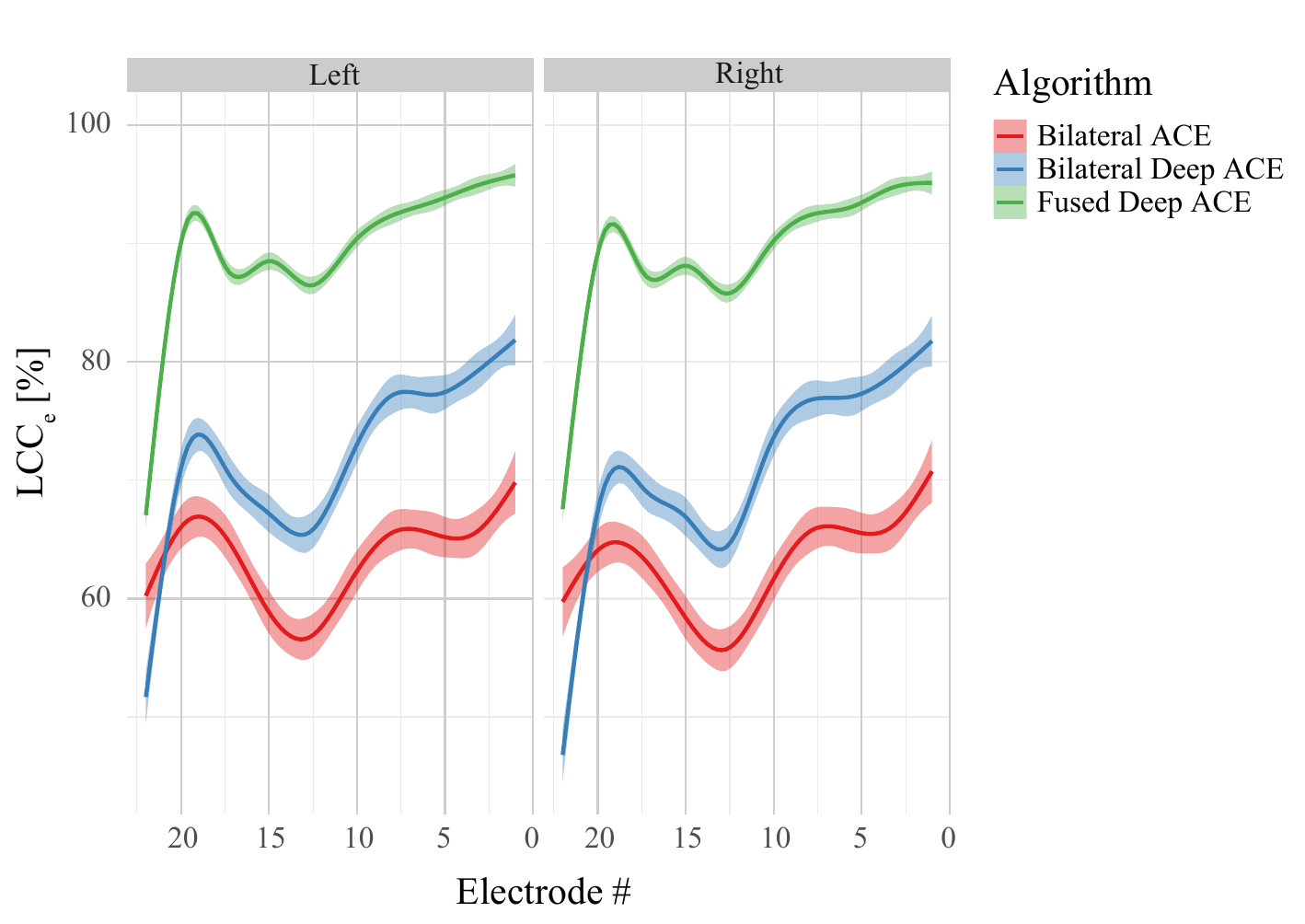}  
    \caption{Polynomial regressions showing the channel-wise LCCs between processed and clean electrodograms for the different algorithms, noises, and listening sides using the HSM dataset. Shaded areas represent the 95\% confidence level interval \cite{rstudio}. Higher electrode numbers represent lower frequencies.}
    \label{lcc_electrodes}
\end{figure}

\begin{figure}[h!]
	\centering
	\includegraphics[width = .48\textwidth]{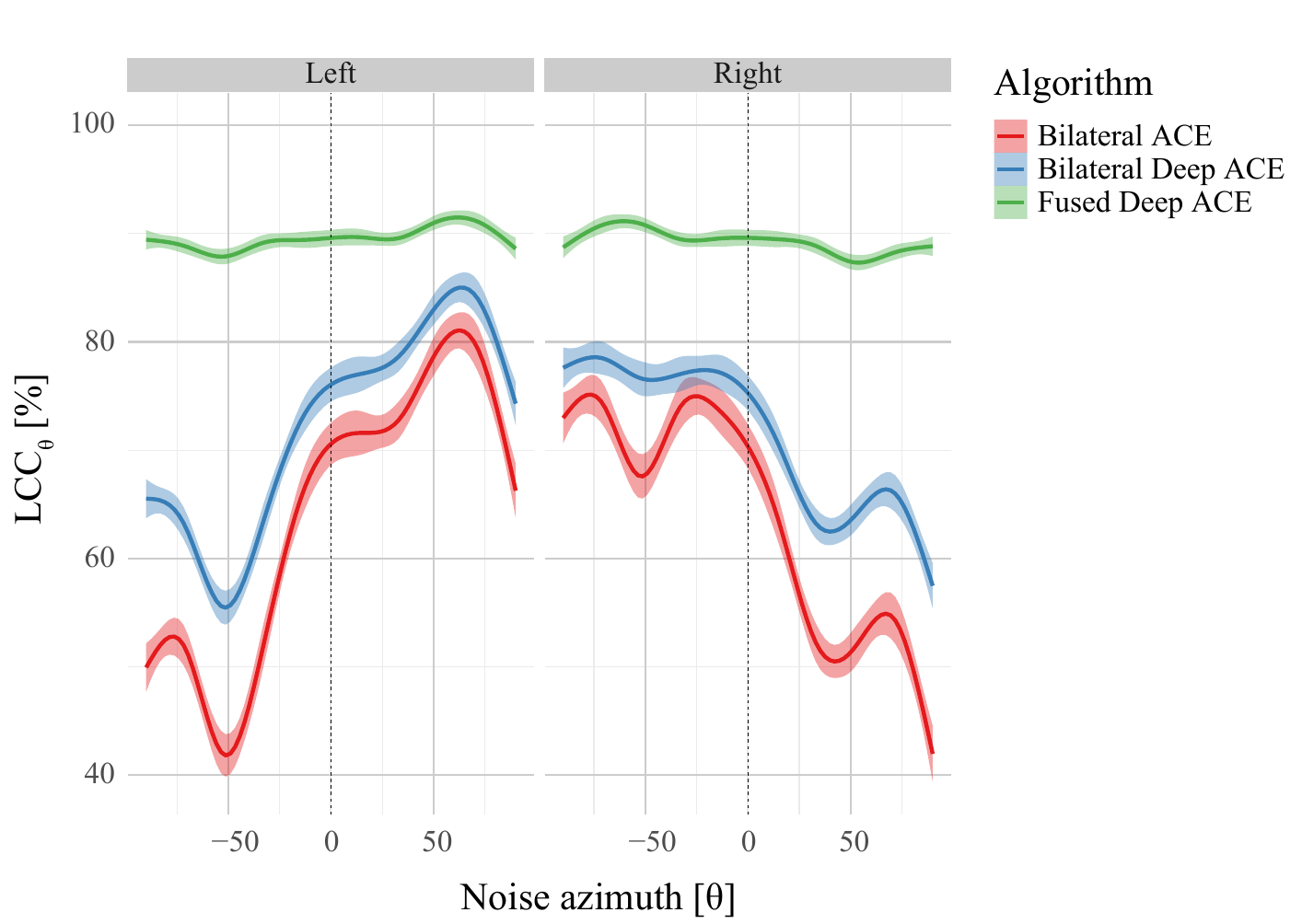}  
    \caption{Polynomial regressions showing the linear cross-correlations between processed and clean electrodograms for the different algorithms averaged across electrodes as a function of the azimuth, noises, and listening sides using the HSM dataset. Shaded areas represent the 95\% confidence level interval \cite{rstudio}. Higher electrode numbers represent lower frequencies.}
    \label{lcc_azimuth}
\end{figure}

\paragraph{\textbf{Linear cross-correlation}}
Figure \ref{lcc_electrodes} illustrates the calculated LCCs across CI electrode numbers for each listing side (averaged across various noise conditions). The data reveals that the fused Deep ACE model exhibits superior performance in terms of channel-wise LCCs. Furthermore, the bilateral Deep ACE model falls between the bilateral ACE and fused Deep ACE algorithms, indicating that the fusion operation contributes to the enhancement of speech in BiCI listening.
Figure \ref{lcc_azimuth} depicts the computed linear cross-correlations with respect to the noise azimuth, considering an average across all electrodes. The data showcases a trend where the LCCs decrease for the bilateral ACE and Deep ACE conditions when the noise source is located ipsilaterally to the CI processor. In contrast, for the fused Deep ACE model, the LCCs appear to remain relatively constant regardless of the azimuth of the interfering noise signal. This observation suggests that the fused Deep ACE model effectively utilizes the fusion operation by leveraging redundant information present on both listening sides.

\begin{figure}[h!]
	\centering
	\includegraphics[width = .48\textwidth]{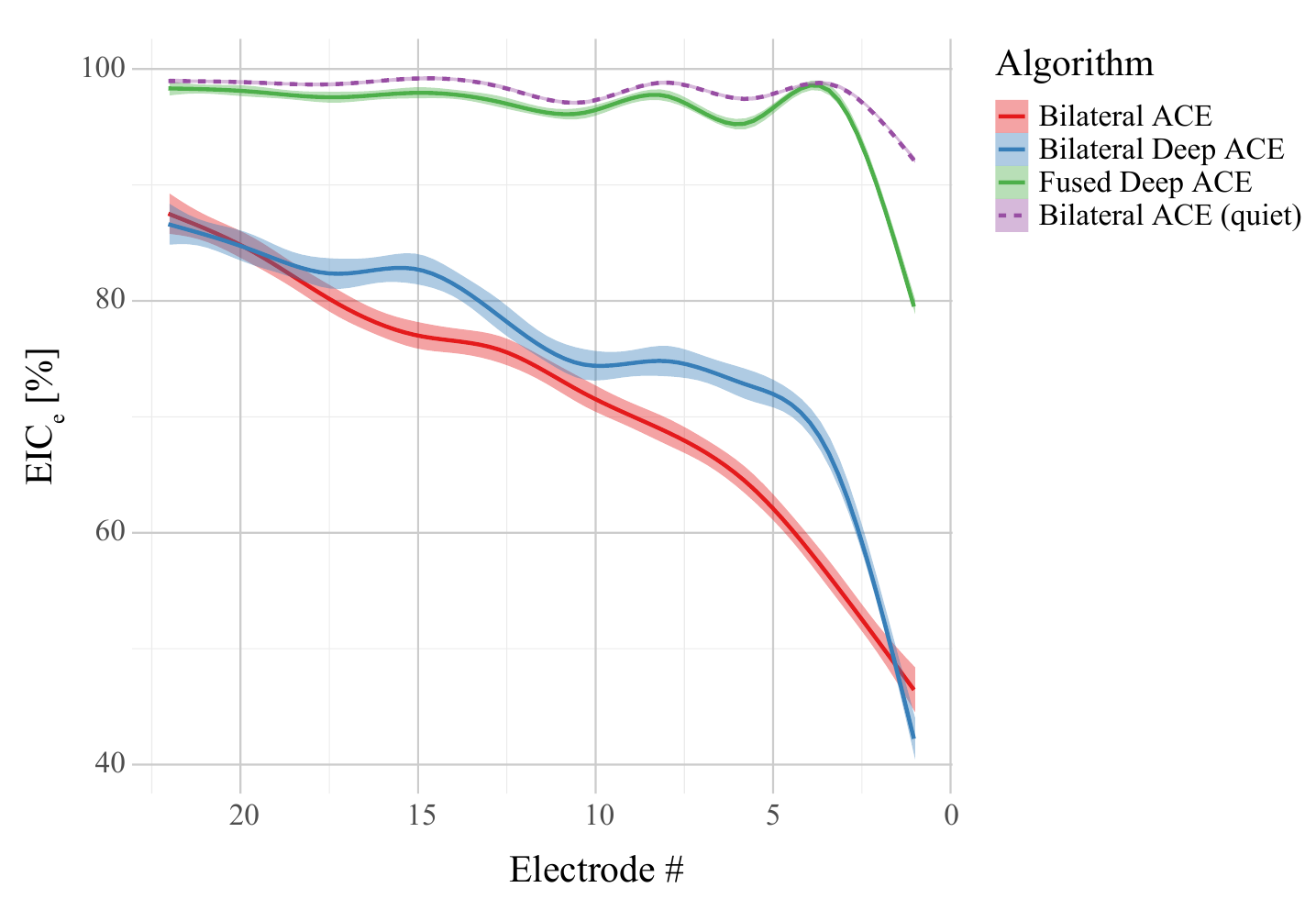}  
    \caption{Polynomial regressions showing the electric interaural coherence (EIC) for each electrode pair averaged across noises and listening sides using the HSM dataset. Shaded areas represent the 95\% confidence level interval \cite{rstudio}. Higher electrode numbers represent lower frequencies.}
    \label{eic_electrodes}
\end{figure}

\begin{figure}[h!]
	\centering
	\includegraphics[width = .48\textwidth]{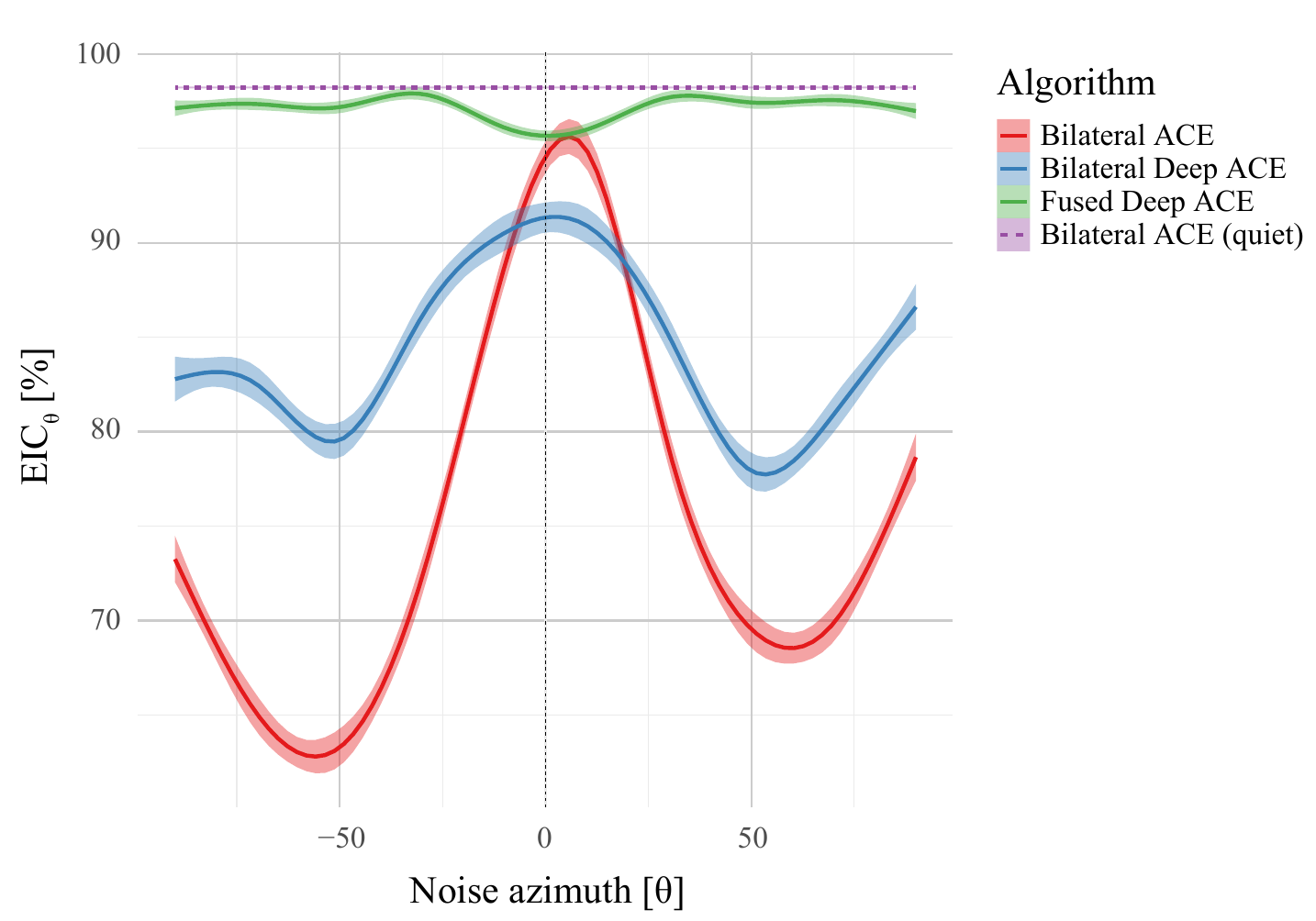}  
    \caption{Polynomial regressions showing the electric interaural coherence (EIC) for each azimuth averaged across electrodes, noises, and listening sides using the HSM dataset. Shaded areas represent the 95\% confidence level interval \cite{rstudio}. Higher electrode numbers represent lower frequencies.}
    \label{eic_azimuth}
\end{figure}

\paragraph{\textbf{Electric interaural coherence}}
Figure \ref{eic_electrodes} visually represents the calculated EIC as a function of the CI electrode numbers for each listening side, with the data averaged across various noise conditions. The results clearly demonstrate that the fused Deep ACE model surpasses the other models in terms of channel-wise EIC. Additionally, the bilateral Deep ACE model occupies an intermediate position between the bilateral ACE and fused Deep ACE algorithms, suggesting that the fusion operation plays a vital role in maintaining the integrity of the speech signal across all frequencies.
In Figure \ref{eic_azimuth}, the calculated EIC  data unveils notable trends based on the noise azimuth. The fused deep denoising model consistently maintains speech correlation, regardless of the noise source's location, showcasing its capacity to sustain speech intelligibility across various noise scenarios. Conversely, the unprocessed condition exhibits higher coherence when the noise originates from the listener's front. However, a shift occurs with the bilateral Deep ACE model, which displays greater coherence when noise is in front but reverses this trend when the noise source widens to azimuths beyond 25 degrees. This pattern suggests that the bilateral Deep ACE model may have limitations in handling denoising when target and interfering signals are co-located.

\subsection{Behavioral results}
\begin{figure}
	\centering
	\includegraphics[width = .48\textwidth]{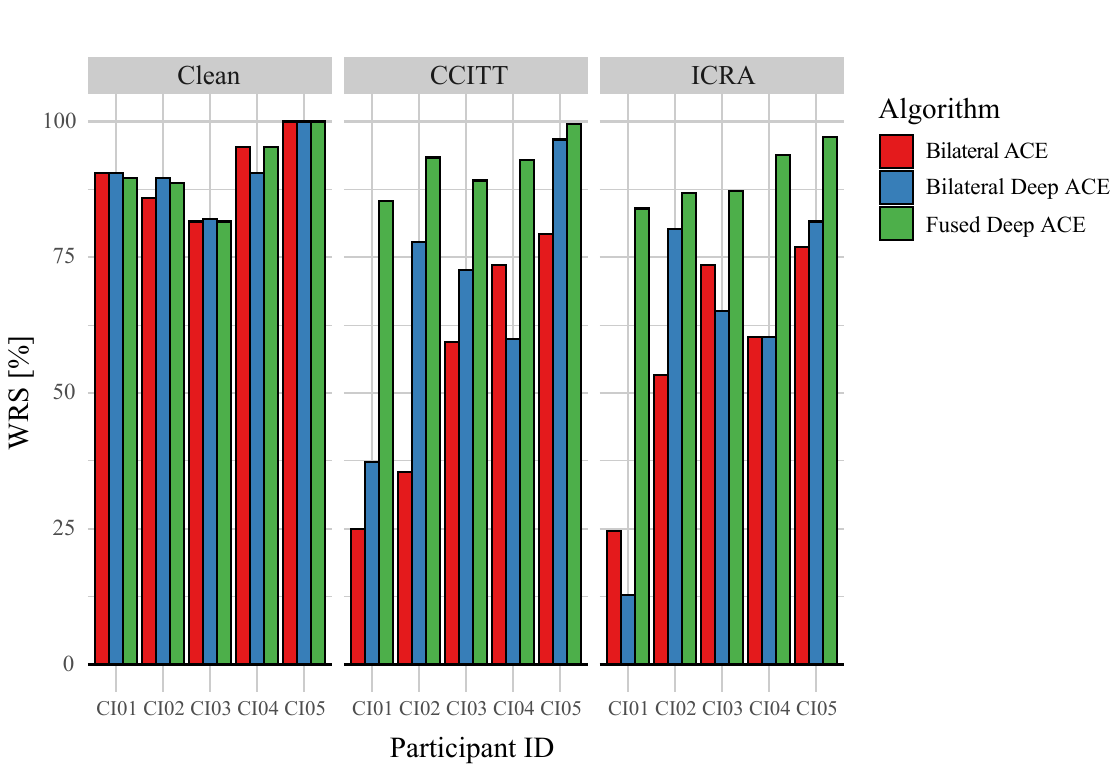}  
    \caption{Bar plots showing the mean individual word recognition scores by subject for the HSM sentence test under CCITT (left panel) and ICRA7 (right panel) noises for all tested algorithms.}
    \label{hsm_ind}
\end{figure}

\paragraph{\textbf{Speech intelligibility}}
Figure \ref{hsm_ind} shows the bar plots of the individual WRS obtained by each of the tested BiCI listeners for the three tested conditions (i.e., clean, CCITT, and ICRA7). The tested SNR for each individual and noise type is shown in Table \ref{Participants}.

\begin{figure}
	\centering
	\includegraphics[width = .48\textwidth]{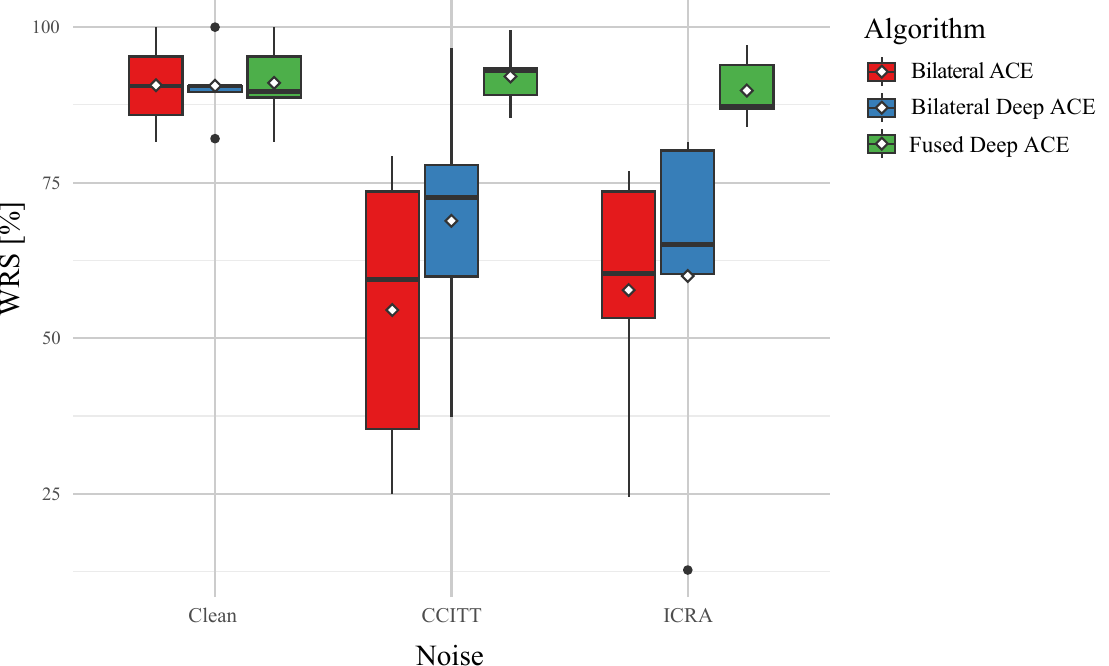}  
    \caption{Box plots of the group word recognition score measured in the five tested BiCI subjects for the three noise conditions. The black horizontal bars within each box represent the median for each condition, the diamond-shaped marks indicate the mean, and the top and bottom extremes of the boxes indicate the $Q_3=75\%$ and $Q_1 = 25\%$ quartiles, respectively. The box length is given by the interquartile range (IQR), used to define the whiskers that show the variability of the data above the upper and lower quartiles (the upper whisker is given by $Q_{3} + 1.5\cdot$IQR and the lower whisker is given by $Q_{1} - 1.5\cdot$IQR \cite{rstudio}).}
    \label{hsm_mean}
\end{figure}

Figure \ref{hsm_mean} displays box plots illustrating the mean WRS measured in the five BiCI subjects across three noise conditions: clean, CCITT, and ICRA7. A Kruskal-Wallis test did not reveal any significant differences in mean speech intelligibility scores for the clean condition ($H(2)=0.04$, $p = 0.98$). However, in the case of the CCITT noisy condition ($H(2)=7.46$, $p = 0.02$) and the ICRA noisy condition ($H(2)=9.57$, $p = 0.008$), the subsequent non-parametric Kruskal-Wallis tests did detect significant differences.

Subsequent pairwise comparisons, conducted using Wilcoxon signed-rank tests, indicated a significant distinction between the unprocessed ($M = 54.52\%$, $SD = 23.65\%$) and the fused deep ACE condition for the CCITT noise ($M = 92.07$, $SD=5.27\%$) conditions ($p = 0.008$). Similarly, significant differences were observed between the unprocessed condition ($M = 57.74\%$, $SD = 20.90\%$) and the fused Deep ACE condition ($M = 89.81\%$, $SD = 5.49\%$) in the ICRA7 noise condition ($p = 0.008$). Additionally, in the ICRA7 noise condition, significant differences were found between the bilateral Deep ACE condition ($M = 60\%$, $SD = 28\%$) and the fused Deep ACE condition ($p = 0.008$).
\paragraph{\textbf{MUSHRA}}

\begin{figure}
	\centering
	\includegraphics[width = .48\textwidth]{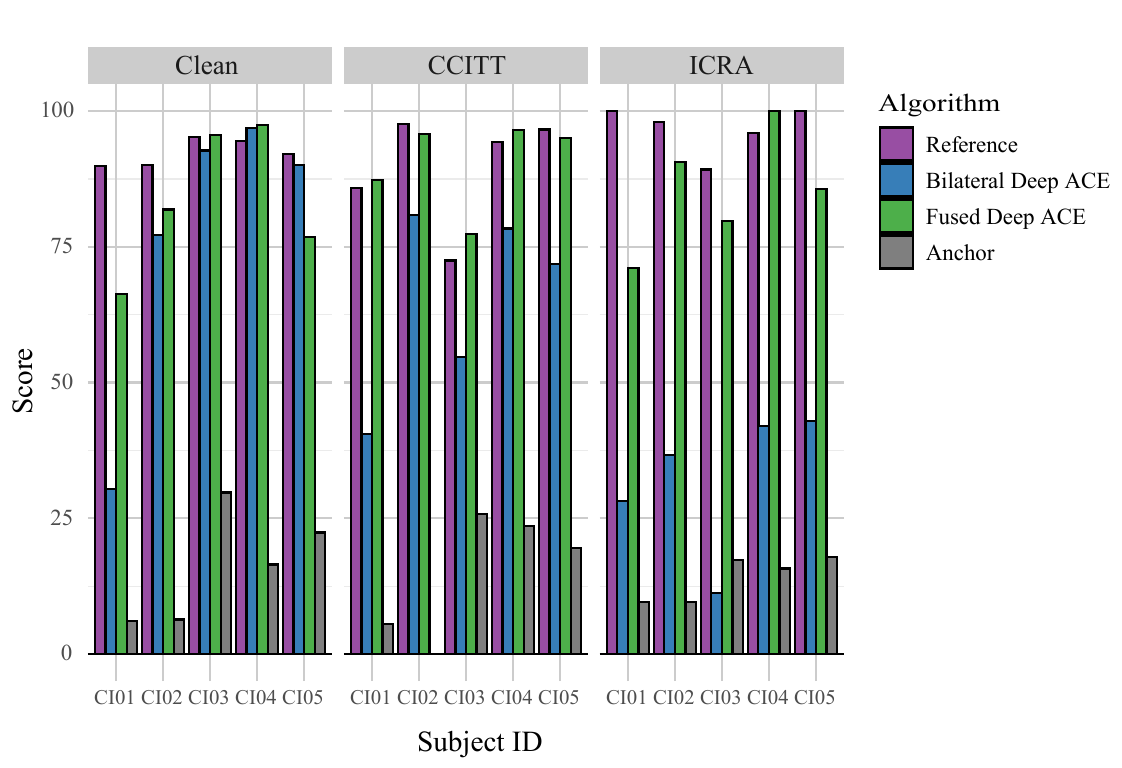}  
    \caption{Bar plots showing the mean individual MUSHRA scores for the HSM sentence test in quiet (left panel), in CCITT noise (center panel), and ICRA7 noise (right panel) noises for all tested algorithms.}
    \label{mushra_ind}
\end{figure}

\begin{figure}
	\centering
	\includegraphics[width = .48\textwidth]{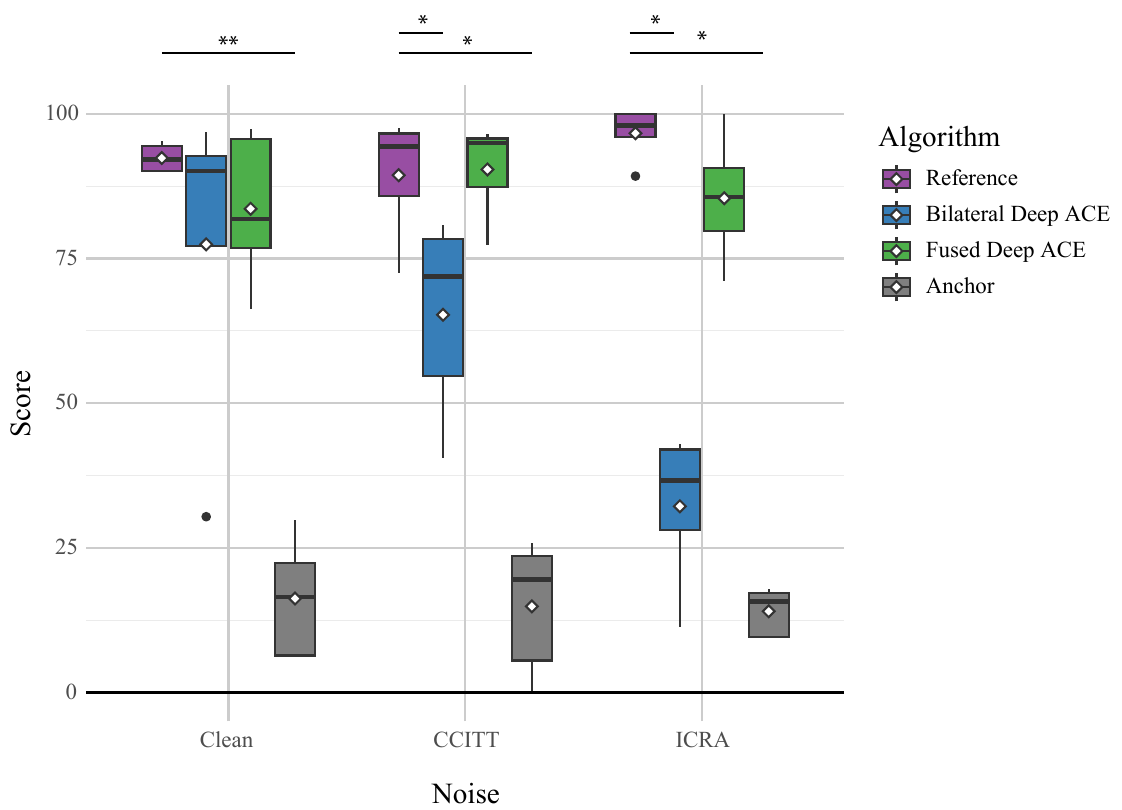}  
    \caption{Box plots of the group MUSHRA score measured in the five tested BiCI subjects for the three noise conditions. The black horizontal bars within each box represent the median for each condition, the diamond-shaped marks indicate the mean, and the top and bottom extremes of the boxes indicate the $Q_3=75\%$ and $Q_1 = 25\%$ quartiles, respectively. The box length is given by the interquartile range (IQR), used to define the whiskers that show the variability of the data above the upper and lower quartiles (the upper whisker is given by $Q_{3} + 1.5\cdot$IQR and the lower whisker is given by $Q_{1} - 1.5\cdot$IQR \cite{rstudio}).}
    \label{mushra_mean}
\end{figure}

Figure \ref{mushra_ind} shows the bar plots of the individual MUSHRA scores obtained by each of the tested BiCI listeners for the three tested noise conditions (i.e., clean, CCITT, and ICRA7).
Figure \ref{mushra_mean} illustrates box plots depicting the group MUSHRA scores obtained from five BiCI subjects under three distinct noise conditions: clean, CCITT, and ICRA7. Three separate Kruskal-Wallis tests, one for each noise condition, unveiled significant differences in the mean MUSHRA scores. Specifically, there were significant differences observed in the quiet condition ($H(3) = 10.99$, $p = 0.01$), the CCITT noise condition ($H(3) = 14.5$, $p = 0.002$), and the ICRA7 noise condition ($H(3) = 16.02$, $p = 0.001$).

Subsequent non-parametric Wilcoxon signed-rank pairwise comparisons further elucidated these differences. In the clean condition, the reference ($M = 92.37$, $SD = 2.46$) obtained higher scores compared to the anchor ($M = 16.23$, $SD = 10.25$) conditions ($p=0.008$). In the CCITT noise condition, the reference ($M = 89.4$, $SD = 10.52$) received higher ratings than both the bilateral Deep ACE ($M = 65.28$, $SD = 17.2$; $p = 0.03$) and anchor ($M = 14.88$, $SD = 11.46$; $p = 0.01$) conditions. Finally, in the ICRA7 noise condition, the reference also achieved higher scores ($M = 96.65$, $SD = 4.46$) compared to the bilateral Deep ACE ($M = 32.17$, $SD = 13.09$; $p = 0.01$) and anchor ($M = 14.02$, $SD = 4.1$; $p = 0.01$) conditions.
\section{Discussion}
In this study, we introduce and evaluate a novel deep learning-based strategy for sound coding in BiCIs. Our approach involves the integration of two monaural end-to-end deep denoising CI sound coding methods through fusion layers that facilitate the exchange of information between the listening sides. This exchange is achieved by combining specific latent representations generated in each monaural model. The presented fused Deep ACE model aims to replicate the ACE sound coding strategy while automatically eliminating unwanted interfering noise from the target speech, all while maintaining minimal processing latency. To be precise, this model introduces the same 2ms latency as the bilateral ACE setup, allowing for potential real-time application of the proposed approach. It's important to note that the transmission of the latent representation must also be taken into account, necessitating efficient coding schemes for the latent space to enable the functional use of the fused Deep ACE model.

Initially, we assess the impact of fusion (fused Deep ACE) by comparing the effectiveness of speech denoising and performance with the bilateral version (bilateral Deep ACE). Furthermore, we compare our approach with the standard clinical BiCI setup, which does not incorporate any denoising (bilateral ACE). Our evaluation involves the testing of this method on speech and the assessment of speech enhancement quality in five BiCI users.

The objective instrumental measures reveal that in quiet environments, there are no discernible differences in speech intelligibility between the bilateral ACE setup, bilateral Deep ACE, and fused Deep ACE models (as shown in Figure \ref{mbstoi_clean}). However, in the context of speech denoising, both bilateral Deep ACE and fused Deep ACE models exhibit improvements in SNR, with the fused Deep ACE model achieving the highest. Surprisingly, the bilateral Deep ACE model performs less effectively when exposed to background ICRA7-modulated noise. This outcome is unexpected, given previous research indicating better results in a similar scenario (as reported in \cite{gajecki2023deep}). This discrepancy could be attributed to the lower SNR used in the current study. Nevertheless, both fused Deep ACE and bilateral Deep ACE models consistently outperform the unprocessed setup in terms of predicted speech intelligibility across all input SNRs.

To assess the extent of clean speech preservation after denoising, we employ objective measures, such as cross-channel and cross-noise azimuths' LCCs. These measures demonstrate that the fused Deep ACE model surpasses the bilateral Deep ACE model in terms of speech-denoising effectiveness and introduces fewer artifacts. This improvement is likely associated with the fused model's ability to exploit the redundancy of speech information shared between sides through the fusion layers. This result is consistent across channels and azimuths. Additionally, as expected, the bilateral Deep ACE model generally exhibits higher LCCs than the unprocessed condition, considering that the unprocessed signal retains all the original interfering noise. It is noteworthy that there is an asymmetry in the LCCs observed in both the bilateral Deep ACE and bilateral ACE conditions (as depicted in Figure \ref{lcc_azimuth}), with lower LCCs measured on the side ipsilateral to the noise source. This asymmetry is also present in the fused Deep ACE model, but to a lesser extent, possibly due to the sharing of speech information between sides.

In the context of BiCI listening, it is crucial to evaluate the retention of EIC after speech denoising, as low EIC has been shown to negatively affect speech intelligibility in BiCI users, as highlighted in \cite{cleary2023effect}. Our assessment reveals that the fused layer achieves the highest EIC scores when measured across azimuths and across electrodes, outperforming the bilateral Deep ACE and bilateral ACE conditions. When observing EIC as a function of the azimuth (as shown in Figure \ref{eic_azimuth}), all three conditions exhibit the highest EIC when speech and noise sources are co-located, aligning with expectations. In this scenario, the unprocessed condition achieves EIC scores closer to those of the fused Deep ACE model, surpassing the scores of the bilateral Deep ACE model. Shows that enhancing speech becomes easier even for the investigated models when interfering noise and target speech are spatially separated, potentially linked to the binaural unmasking phenomenon observed in human binaural hearing. This underscores the significance of higher SNR listening sides in BiCI speech denoising, particularly when speech information is shared between sides, as facilitated by the fusion layers in our approach.

The behavioral results in quiet conditions reveal no significant differences in speech intelligibility among the ACE, bilateral Deep ACE condition, and fused Deep ACE sound coding strategies, corroborating the findings from objective measures. This consistency is further confirmed by the MUSHRA test, where no discrepancies in scores are observed among the reference, bilateral Deep ACE, and fused Deep ACE conditions. However, in noisy speech conditions, speech intelligibility experiments showed that the fused Deep ACE model outperforms the bilateral ACE and bilateral Deep ACE conditions, while the bilateral Deep ACE condition surpasses ACE only in the presence of CCITT noise, failing to yield improvement when ICRA7 background noise is present. These results align with the observed SNR improvements in these conditions.

Furthermore, the MUSHRA scores indicate that all BiCI users were capable of identifying the reference and the anchor. In terms of denoising algorithms, the scores were consistently lower for the bilateral Deep ACE model compared to the reference, particularly for both CCITT and ICRA7 conditions. This concurs with the measured speech intelligibility results, implying that speech intelligibility may be significantly affected not only by the limited SNR improvement in this condition but also by the bilateral distortions introduced by the bilateral Deep ACE  model.

\section{Conclusion}
This study underscores the potential of a fused deep learning-based BiCI sound coding strategy (fused Deep ACE) in enhancing speech, especially when speech and interfering noise sources are spatially separated. Notably, the approach's ability to retain interaural coherence compared to the bilateral Deep ACE  model is highlighted. The proposed fused Deep ACE model achieved significant improvement in objective instrumental measures as well as in the listening experiments with BiCI participants. However, it is crucial to recognize that this approach may not be optimal in all listening conditions, as it may compromise binaural and spatial awareness, akin to the effects of front-end beamformers. Further research is warranted to strike a balance between achieving high speech denoising performance and maintaining spatial awareness through fusion layers, which may entail a trade-off, as outlined in \cite{marquardt2015interaural}. Nevertheless, our presented approach exhibits promising speech-denoising performance and may prove beneficial in specific listening conditions.

\end{document}